\documentclass[%
 reprint,
 amsmath,amssymb,
 aps,
]{revtex4-2}

\usepackage{graphicx}
\usepackage{dcolumn}
\usepackage{bm}
\usepackage{relsize}
\usepackage{amsmath}
\usepackage{upquote}
\usepackage[export]{adjustbox}

\begin{document}

\preprint{APS/123-QED}

\title{Field-driven cluster formation in two-dimensional colloidal binary mixtures}

\author{Dingwen Qian}

\affiliation{ 
Applied Physics Program, Northwestern University,  Evanston, Illinois 60208, USA
}%
\author{Monica Olvera de la Cruz}%
 \email{m-olvera@northwestern.edu}
\affiliation{ 
Applied Physics Program,  Departments of Materials Science and Engineering, Chemistry,  and Physic and Astronomy,  Northwestern University, Evanston, Illinois 60208, USA
}%

\date{\today}

\begin{abstract}

We study size- and charge-asymmetric oppositely charged colloids driven by an external electric field. The large particles are connected by harmonic springs, forming a hexagonal-lattice network while the small particles are free of bonds and exhibit fluid-like motion. We show that this model
exhibits a cluster formation pattern when the external driving force exceeds a critical value. The clustering is accompanied with stable wavepackets in vibrational motions of the large particles.

\end{abstract}

\maketitle


\section{\label{sec:intro}Introduction}
In solids, the atoms vibrate around their equilibrium positions and the excitations of vibrational motions, phonons, are well described by elastic theory \cite{Chen2013}. Although particles in fluids have vibrational motions, they do not have well-defined equilibrium positions. Instead of tracking the motion of individual fluid particles, it is more appropriate to describe them with continuous quantities, such as density, velocity, and pressure. Active systems can be solid or fluid. While active solids have well-defined reference positions, active fluids do not. Most well-studied active systems are fluids \cite{Loi2008, Toner1998, Rubenstein2014}, however, interest in active solids has grown recently \cite{Baconnier2022, Ferrante2013}.

In many physical systems, people have observed the coexistence of vibrational and fluid-like degrees of freedom. Examples include ion cores and conducting electrons in metals, rigid and mobile ions in superionic conductors \cite{Boyce1979} and most recently, superionic behavior observed in size-asymmetric colloidal compounds \cite{Leunissen2005, Girard2019, Wang2022, Ehlen2021, Lopez-Rios2021, Lin2020, Lin2022, Padilla2021}. While there are many reports on the equilibrium behavior of these mixtures, the far-from-equilibrium behavior is still not studied. In this paper, we explore the far-from-equilibrium behavior of systems with both \textit{vibrational} and \textit{fluid-like} degrees of freedom. 

Here, we study size- and charge-asymmetric oppositely charged colloidal compounds under an external electric field in a two-dimensional space. The study of two-dimensional colloidal mixtures is highly relevant to the behavior of materials confined at interfaces in both equilibrium \cite{Thorneywork2017, Keim2004, Bernard2011, Bian2021} and nonequilibrium \cite{Zottl2014, Glanz2012} conditions. When colloidal particles sediment onto the base of a glass sample cell, they form a monolayer of colloidal mixtures and their motions are in the two directions parallel to the glass base \cite{Thorneywork2017}. Recent advancement in the synthesis of colloidal disks \cite{Qu2021} inspires us to study size-asymmetric colloidal disks in a two-dimensional space. A particular feature of our model systems is harmonic interactions between the large particles. The large colloidal particles are connected by harmonic springs and form a hexagonal lattice. The harmonic springs enforce elastic coupling between the motion of the large particles in the far-from-equilibrium regime, and elasticity plays an important role in the phenomena we describe in this paper. Experimentally, the development of programmable DNA origami nanosprings makes it possible to create springs with adjustable lengths and strengths \cite{Wickham2016}. Two-dimensional networks of polymer-linked nanoparticles have been synthesized \cite{Hu2022}, which is similar to the network of large colloidal particles we propose in this paper. Another important feature of our systems is an extremely large size ratio. The size of the small particles is set to be much smaller than the large particles so that the small particles drift across the lattice under the external field, showing the behavior of a continuous flow. Under external driving force, the small particles flow through the lattice formed by the large particles, resembling the flow of fluid through the interstitial spaces in a foam \cite{Zhu2021, Koehler2000}. In this study, in analogy to the solitons found in the flow of fluid through foams, we observe cluster formation of small particles and soliton-like vibrational motions of the large particles when the external driving force exceeds a critical value. The soliton-like vibrational motions mean that wavepackets of displacement vectors of the large particles travel with stable shapes.

\section{\label{sec:Model}Model and methodology}

\begin{figure}
  \includegraphics[width=\linewidth]{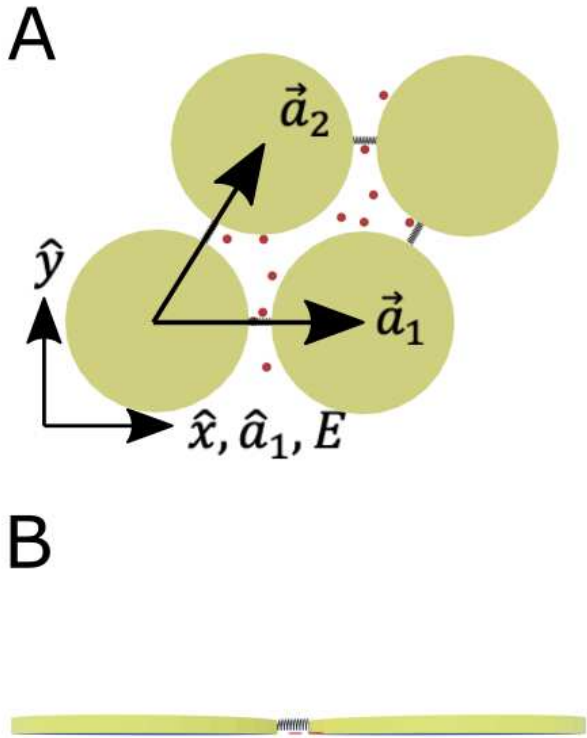}
  \caption{\label{fig:sketch} Sketches of the system geometry. (A) Top view of one unit cell of the hexagonal lattice. The lattice vectors are $\Vec{a}_1=[45r,0]$ and $\Vec{a}_2=[\frac{45r}{2},\frac{45\sqrt{3}r}{2}]$. The electric field is applied in the direction of $\Vec{a}_1$, which is the x-direction of the simulation box. (B) Side view of the 3D structures of the large colloidal particles, small colloidal particles, and nanosprings connecting the large particles. The height of the small particles is much lower than the height of the large particles, $h\ll H$, and the monomers of the nanosprings are typically at the length scale of $1\ \mathrm{nm}$. Therefore, the nanosprings are not an obstacle for the small particles.}
\end{figure}

Here we briefly summarize the system design and geometry. The motions of the size-asymmetric colloidal disks are confined to the two directions parallel to the substrate. The size ratio between the large and small colloidal particles is $R:r=20:1$ and the number ratio between the large and small particles is $N:n=1:12$. The radius of the small particles is $r=50\ \mathrm{nm}$, which is the unit length scale in our system. The large colloidal particles are connected by harmonic springs, forming a hexagonal lattice with lattice constant $a=45r$. FIG. \ref{fig:sketch} shows the geometry of one unit cell of the system from the top view and the side view. The total area fraction of the colloidal particles is $\phi=0.738$. The simulations are conducted in a square box containing $60\times 60$ unit cells with periodic boundary conditions. The particle number, volume, and temperature are constant.

We introduce here the interactions and external driven forces in our Brownian dynamics simulations. Since the system is in a low Reynolds number regime, overdamped Langevin dynamics is implemented to simulate the equilibrium and nonequilibrium behavior of our system, which neglects the inertia of particles. We use the LAMMPS package to conduct the Brownian dynamics simulations \cite{Thompson2022}. We do not consider hydrodynamics effects here.
The equation of motion for particle i is:
\begin{equation}
    \frac{\mathrm{d}\textbf{r}_{i}}{\mathrm{d}t}=\frac{-{\boldsymbol\nabla}_{i} U+\textbf{F}^i_{ex}}{\gamma_{i}}+ \boldsymbol\xi_{i}
\end{equation}
where $\textbf{r}_{i}$ is the position vector of the particle $i$, $U$ is the total conservative force potential energy,  $\textbf{F}^i_{ex}=\textbf{E}q_{i}$ is the external force acting on the particle $i$, and $\gamma_{i}$ is the drag coefficient of particle $i$. The charge of the large particles is $q_L=96\ e$ and the charge of the small particles is $q_S=-8\ e$. The charge ratio between the large and small particles is $q_L:q_S=12:-1$, which respects the charge neutrality considering the number ratio is $N:n=1:12$. The drag force on particle $i$ depends on the drag coefficient and velocity of the particle $\textbf{F}_{Di}=-\gamma_{i}\textbf{v}_i$. Here we explain how we set the drag coefficients of the colloidal disks. We set the height of the colloidal disks to be much smaller than their radius, i.e., $h=5\ \mathrm{nm}\ll r$ and $H=50\ \mathrm{nm}\ll R$. In a low Reynolds number fluid with viscosity $\eta$, the drag force on a disk with radius $r$, negligible height $h\ll r$ and moving in the plane with speed $\boldsymbol{v}$ has an exact formula $\boldsymbol{F}_D=-\frac{32}{3}\eta r\boldsymbol{v}$, which is independent of the thickness of the disks [28]. Therefore, the friction coefficients of the small particles and large particles are $\gamma_{S}=\frac{32}{3}\eta r$ and $\gamma_{L}=\frac{32}{3}\eta R$ respectively, where the water viscosity is $\eta=1\times10^{-3}\ \mathrm{Pa\cdot s}$. It follows that the friction coefficient for the large particles is 20 times the friction coefficient for the small particles, $\gamma_{L}=20\gamma_{S}$. The thermal force $\boldsymbol\xi_{i}$ is a white noise with zero mean, satisfying $\langle\textbf{$\xi$}_{i{\alpha}}\textbf{$\xi$}_{j{\beta}}\rangle={2\frac{k_{B}T}{\gamma_{i}}{\delta_{ij}}{\delta_{{\alpha}{\beta}}}{\delta}(t-t')}$, where $i$, $j$ denote the particle indices and  $\alpha$, $\beta$ denote the spatial directions. $k_{B}T$ is the thermal energy at $T=298\ \mathrm{K}$ and set as the unit energy scale in the simulation. The Brownian time of the small particles $\tau=\frac{r^{2}}{D}=3.3\times 10^{-4}\ \mathrm{s}$ is set as the time unit, where $D=\frac{k_{B}T}{\gamma_{S}}$ is the bare diffusion coefficient of the small particle. The timestep for integration is $\Delta t=10^{-4}\tau$. 

The potential energy of total conservative interactions $U$ consists of bonded and non-bonded interactions. The bonded interaction is the harmonic potential between neighboring large particles, $U_{spring}=\mathlarger{\mathlarger{\sum}}_{\{ij\}(nn)}\frac{1}{2}K(r_{ij}-a)^2$, where $(nn)$ denotes nearest neighbors, $K$ is the spring constant, $r_{ij}$ is the distance between the neighboring large particles and $a$ is the equilibrium length of the harmonic springs, which equals the lattice constant. In this report, we use a dimensionless spring constant to quantify the strength of the springs, $\tilde{K}=\frac{Kr^2}{k_{B}T}$, where $r$ is the radius of the small particles and the unit length scale in our simulations. The non-bonded interactions are composed of two parts, the hardcore interaction and the screened Coulomb interaction.
 We use the repulsive Weeks-Chandler-Andersen (WCA) potential to model the hardcore interaction, $U_{WCA}(r_{ij})$ given by
 \begin{equation}
 \begin{cases}
    
    4\varepsilon\left[\Big(\dfrac{\sigma_{ij}}{r_{ij}}\Big)^{12}-\Big(\dfrac{\sigma_{ij}}{r_{ij}}\Big)^{6}\right]+\varepsilon, & r_{ij}<2^{1/6}\sigma_{ij}.\\
    0, & r_{ij}>2^{1/6}\sigma_{ij}.
  \end{cases}
\end{equation}
and Yukawa potential to model the screened electrostatic interaction,
\begin{equation}
    \beta U_{el}(r_{ij})=\frac{\lambda_{B}q_{i}q_{j}\exp[-{\kappa}(r_{ij}-\sigma_{ij})]}{(1+\kappa R_{i})(1+\kappa R_{j})r_{ij}}
\end{equation}
Here, $\sigma_{ij}=R_{i}+R_{j}$ where the $R_{i}$ and $R_{j}$ are the radii of particle $i$ and $j$ respectively. The WCA interaction strength $\varepsilon=10\ k_{B}T$, $\beta=\frac{1}{k_{B}T}$, and $\lambda_B=\frac{e^2}{4\pi \epsilon_0 \epsilon_r k_{B}T}$ is the Bjerrum length. The Bjerrum length equals $0.7\ \mathrm{nm}$ for water solvent at $T=298\ \mathrm{K}$, and $q_i$ and $q_j$ are the charges of the particle $i$ and $j$ in the unit of the elementary charge respectively. The Debye length $\lambda_D=\kappa^{-1}=(4\pi \lambda_{B} \mathlarger{\mathlarger{\sum}}^N_{i=1}n_i z^2_i)^{-\frac{1}{2}}$ is a measure of how far a charge carrier's electrostatic effects persist in a solution with free ion concentration $n_i$ and ion charge numbers ${z_i}$. The environment is deionized water with ion concentration $n=10^{-7}\ \mathrm{mol/L}$.
The Debye length equals $1\ \mathrm{\mu m}$ in the deionized water at $T=298\ \mathrm{K}$, which is the radius of the large particles. This means the electrostatic interaction persists up to the radius of the large particles. 

Before we conduct the simulations on the nonequilibrium conditions, we prepared the systems by allowing them to reach thermodynamics equilibrium. We set the external force equal to zero and run the simulations for $10^{2}\tau$. The pressure and internal energy of the systems become stable, and we see that as the criterion of equilibrium for the systems. Then, we apply an external force on the systems and let them evolve under nonequilibrium conditions. In this study, we focus on the nonequilibrium steady state (NESS) which is defined when the number of clusters reaches a constant value. In our simulations, we observe that clusters of small particles form and change in sizes until the system reaches this nonequilibrium steady state (NESS). The typical time to reach the nonequilibrium steady state is $10^{4}\tau$. 

The particles are driven by an external electric field in the x-direction. The large particles are driven to move to the right and the small particles are driven to move to the left. From $\textbf{F}^i_{ex}=\textbf{E}q_{i}$, we know that the magnitude of the external driving force on the large particles is 12 times the magnitude of the external driving force on the small particles. To quantify how strong the external driving force is compared to the thermal noise, we use the dimensionless driving force $\tilde f\equiv \frac{f_{ex}r}{k_{B}T}=\frac{Eqr}{k_{B}T}$, where $f_{ex}$ is the magnitude of the external driving force on the small particles, $r$ is the radius of the small particles and $k_{B}T$ is the thermal energy.

The center of mass of all large colloidal particles drifts due to the external field and the collision with the small particles. When we look into the vibrational motions of the large particles, it is more convenient to choose the center of mass of all large particles as the frame of reference, which we call the COM frame in this report. We only use the COM frame when we present the results of the vibrational motions of the large particles, and use the laboratory frame of reference for other parts of the report. The transformation of the coordinates of particles from the laboratory frame of reference to the COM frame is $\tilde{\textbf{r}}_{i}(t)=\textbf{r}_{i}(t)-\textbf{R}_{cm}(t)$, where $\tilde{\textbf{r}}_{i}(t)$ is the position vector of the particle $i$ in the COM frame at time $t$, the $\textbf{r}_{i}(t)$ is the position vector of the particle $i$ in the laboratory frame of reference at time $t$ and $\textbf{R}_{cm}(t)=\frac{1}{N}\mathlarger{\mathlarger{\sum}}_{i\in Large}\textbf{R}_{i}(t)$ is the position vector of the center of mass of all large particles in the laboratory frame of reference at time $t$. We use $\textbf{R}_{i}$ to denote the position vectors of the large particles while using $\textbf{r}_{i}$ to denote the position vectors of any particles. In the COM frame, the displacement vector of one large particle is $\delta \tilde{\textbf{R}}_{i}(t)=\tilde{\textbf{R}}_{i}(t)-\tilde{\textbf{R}}_{i0}$, where $\tilde{\textbf{R}}_{i0}$ is the reference position of the large particle $i$ and does not change with time in the COM frame. It follows that $\mathlarger{\mathlarger{\sum}}_{i\in Large}\delta \tilde{\textbf{R}}_{i}(t)=0$. The displacement vector fields are calculated in microstates (snapshots) of the systems. We also calculate the vibrational spectra of the large particles using the discrete Fourier transform of the time profile of the velocities of the large particles.
\begin{equation}
   \textbf{v}_n(\nu)= \frac{1}{N_{frame}}\sum_{j=0}^{N_{frame}-1}{\textbf{v}_n(t_j)}exp[i2\pi{\nu}t_j]
\end{equation}
\begin{equation}
    I_m(\nu)=\frac{\tau^2}{r^2}\frac{1}{N_{particles}}{\sum_{n=1}^{N_{particles}}{{\lvert}{v}_{nm}(\nu)\rvert}^2}
\end{equation}
The vibrational spectra are calculated from velocites recorded from $N_{step}=1000$ consecutive timeframes. The index $m$ denotes the component of the velocities, which can be $x$ or $y$ in our systems. The vibrational intensity is  dimensionless due to the prefactor $\frac{\tau^2}{r^2}$ .

In this report, we focus on the effects of varying the external driving force and the spring constant of the springs between the large particles. The dimensionless spring constants are $\tilde{K}=20, 30, 40$, corresponding to $K=\frac{\tilde{K}k_{B}T}{r^2}=3.36\times10^{-2},\ 5.04\times10^{-2},\ 6.72\times10^{-2}\ \mathrm{pN/\mu m}$. The strongest dimensionless driving force is $\tilde{f}=80$, corresponding to $E=\frac{\tilde{f}k_{B}T}{qr}=5.13\ \mathrm{mV/nm}$, which is much lower than the breakdown field of water $E_{B}=65\ \mathrm{mV/nm}$ \cite{Jones1994}.

\begin{figure}
  \includegraphics[width=\linewidth]{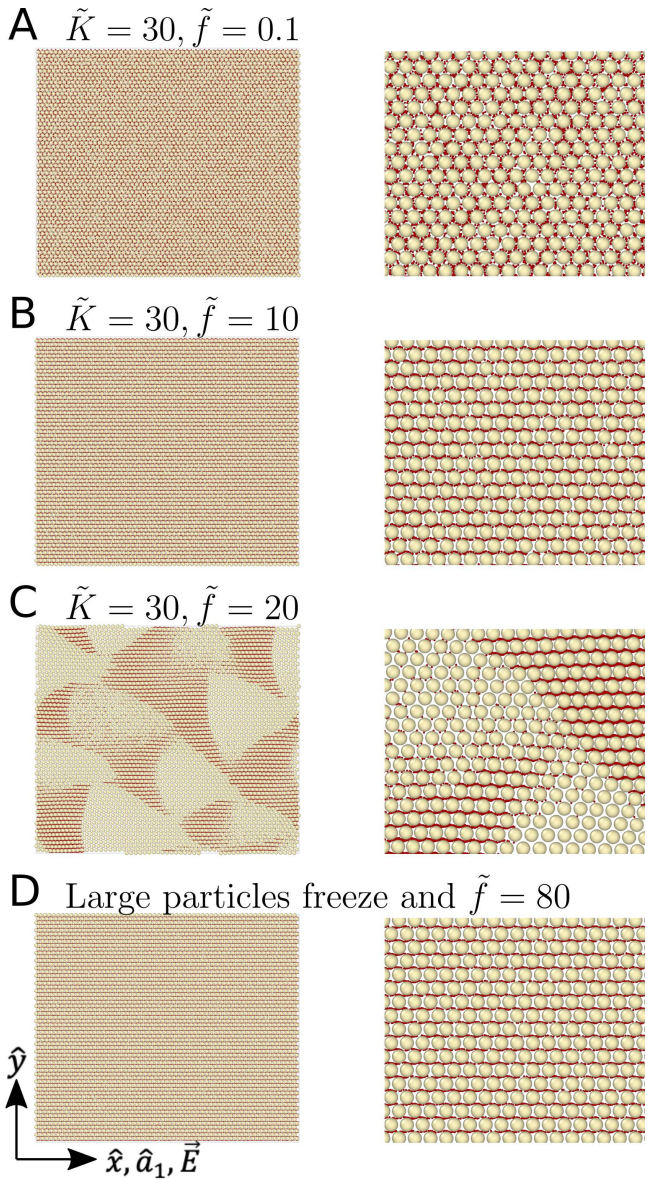}
  \caption{\label{fig:snapshots}Representative snapshots of colloidal mixtures driven in opposite directions with the dimensionless spring constant $\tilde{K}=30$. The left figures represent the whole simulation boxes, while the right figures are enlarged portions of the left figures. All figures are enlarged 4 times from left to right. In addition, the small particles are enlarged 4 times the original size to make them visible. The snapshots are made using the OVITO package \cite{Stukowski2010}. (A) The dimensionless spring constant and driving force are $\tilde{K}=30$ and $\tilde{f}=0.1$, respectively. The driving force is much smaller than the thermal noise, and the small particles distribute homogeneously in the lattice. (B) The dimensionless spring constant and driving force are $\tilde{K}=30$ and $\tilde{f}=10$, respectively. The small particles form lanes in the channels between the large particles, while the large particles vibrate randomly. (C) The dimensionless spring constant and driving force are $\tilde{K}=30$ and $\tilde{f}=20$, respectively. The small particles aggregate into clusters in the channels between the large particles.  (D) The large particles are frozen on their reference positions forming a hexagonal lattice, while the small particles are driven by the dimensionless external force $\tilde{f}=80$. The small particles form lanes in the channels between the large particles. }
\end{figure}
\section{\label{sec:Results}Results and discussion}
\begin{figure}
  \includegraphics[width=\linewidth]{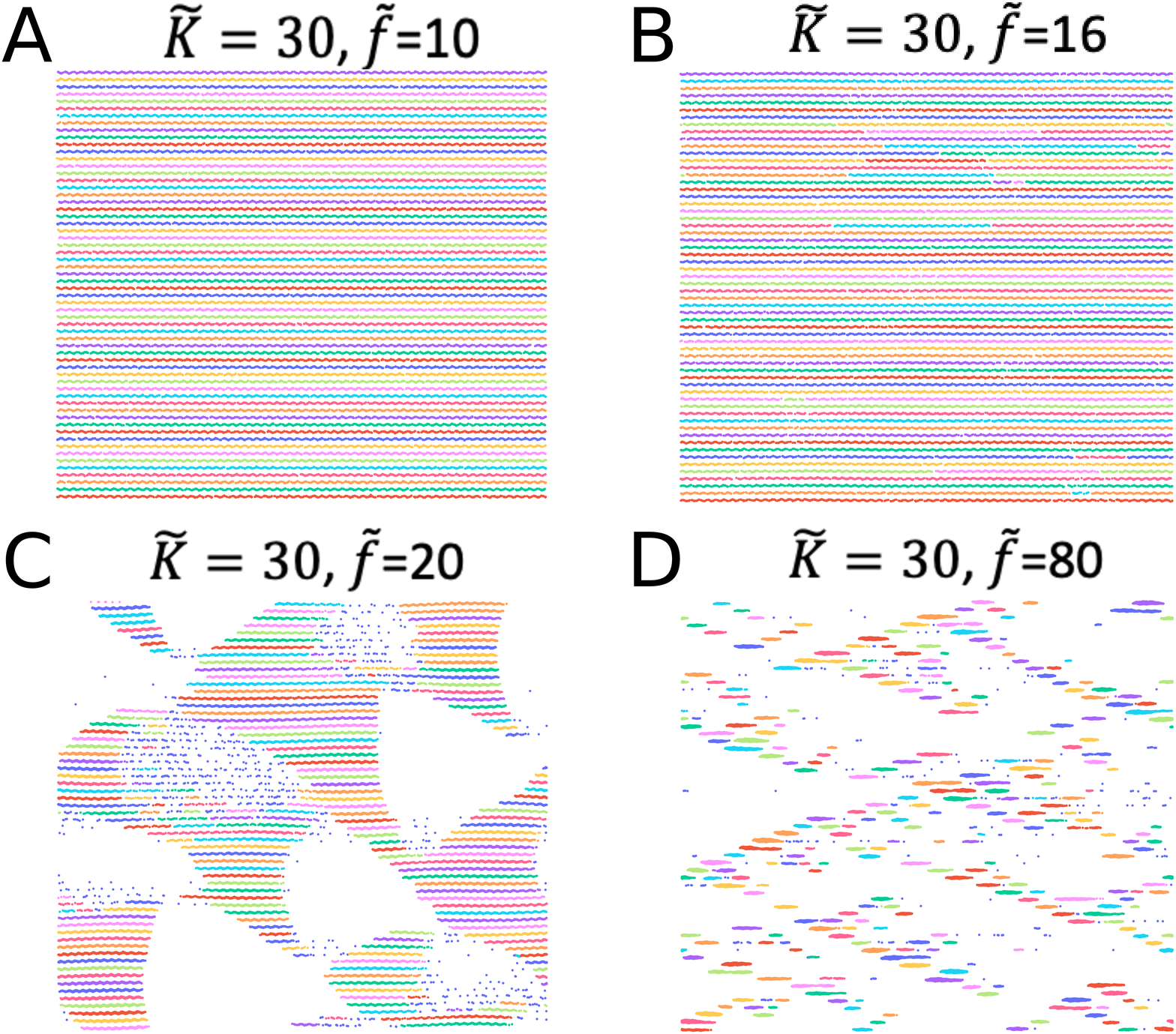}
  \caption{\label{fig:clusters}Illustrations of different states in the driving colloidal mixtures with the dimensionless spring constant $\tilde{K}=30$. Here we show only the small particles, which are divided into clusters. The neighboring clusters are distinguished by their colors. The small particles are enlarged 6 times the original size to make them visible. (A) The dimensionless spring constant and driving force are $\tilde{K}=30$ and $\tilde{f}=10$, respectively. The system is in the lane state, where the small particles form clusters that are percolated only in the x-direction. (B) The dimensionless spring constant and driving force are $\tilde{K}=30$ and $\tilde{f}=16$, respectively. The system is in the intermediate state, where x-percolated clusters and non-percolated clusters coexist. (C) The dimensionless spring constant and driving force are $\tilde{K}=30$ and $\tilde{f}=20$, respectively. The system is in the cluster state, where there are only non-percolated clusters. (D) The dimensionless spring constant and driving force are $\tilde{K}=30$ and $\tilde{f}=80$, respectively. The system is in the cluster state at a higher field, where the clusters are more round in shape. The clusters in the neighbor channels separate from each other in the x-direction with a nearly constant distance. }
\end{figure}

When $\tilde f \ll 1$, the small particles distribute homogeneously in the lattice (FIG. \ref{fig:snapshots}A). When $\tilde f \gg 1$, the small particles form lanes in the channels (FIG. \ref{fig:snapshots}B). The lane formation is observed in size-symmetric colloidal mixtures when the particles are not connected by the springs \cite{Vissers2011,Bagchi2022, Sutterlin2009,Dzubiella2002,Rex2007,Reichhardt2018,Li2021,Klymko2016,Wachtler2016}. In our systems, the small particles form lanes, but the large particles do not aggregate, due to the springs. Interestingly, we observe the formation of clusters of small particles with further increasing $\tilde{f}$ (FIG. \ref{fig:snapshots}C). When we freeze the motion of the large particles, the small particles can still form lanes at high fields, but no cluster formation is observed (FIG. \ref{fig:snapshots}D). This means that in our systems, lane formation does not require the motion of the large particles, while cluster formation only happens when the large particles are allowed to vibrate.

\subsection{\label{sec:Solitary vibration}Lane to cluster transition and phase diagram}

\begin{figure}
  \includegraphics[width=\linewidth]{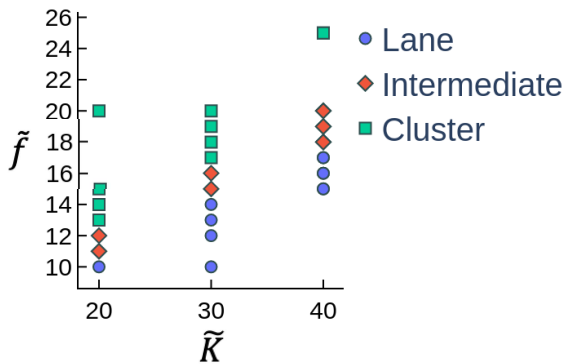}
  \caption{\label{fig:phase diagram}Phase diagram of driven colloidal mixtures. The horizontal axis is the dimensionless spring constant and the vertical axis is the dimensionless external driving force. The phase diagram shows that the transition field increases as the spring constant increases. We only show the data points with external driving forces close to the transition.}
\end{figure}
First of all, we give the definition of a cluster of small particles. Two small particles are considered \textit{neighbors} if they are within the Debye length $\lambda_D$, since they do not interact with each other strongly when they are separated beyond this distance. Two particles are considered in one cluster if we can find a path of \textit{neighbors} connecting them. One cluster is considered percolated in one direction if this cluster goes across the periodic boundary and forms a loop. If we are in the equilibrium state or $\tilde{f} \ll 1$, the small particles distribute homogeneously in the system, therefore they are all connected and considered to be in one cluster. This cluster is percolated in both x and y directions. The lane state occurs when all clusters are percolated in the x-direction and not percolated in the y-direction. The lanes of small particles drift in the channels between the large particles, and the small particles distribute homogeneously inside the channels. Meanwhile, the cluster state occurs when all clusters are not percolated. It is natural to define one intermediate state, where x-percolated clusters and non-percolated clusters coexist.

With these definitions, we find that when we increase the external driving force, the system evolves from the lane state (FIG. \ref{fig:clusters}A) to the intermediate state (FIG. \ref{fig:clusters}B), then into the cluster state (FIG. \ref{fig:clusters}C and D). In the cluster state, the small particles travel in clusters that have constant velocities and stable shapes. As we further increase the external driving force, the shapes of the clusters become more and more round, which can be quantified by asphericity. We will introduce asphericity in the next subsection. We also observe that the clusters in the neighbor channels are repulsive to each other. They separate from each other in the x-direction at a nearly constant distance. The transition driving force is defined as the lowest driving force where the cluster state emerges. When the spring constant increases, we observe that the transition driving force also increases. This is shown in the phase diagram in FIG. \ref{fig:phase diagram}. In the section on the vibrational motions of the large particles, we further discuss this trend.

\begin{figure}
  \includegraphics[width=\linewidth]{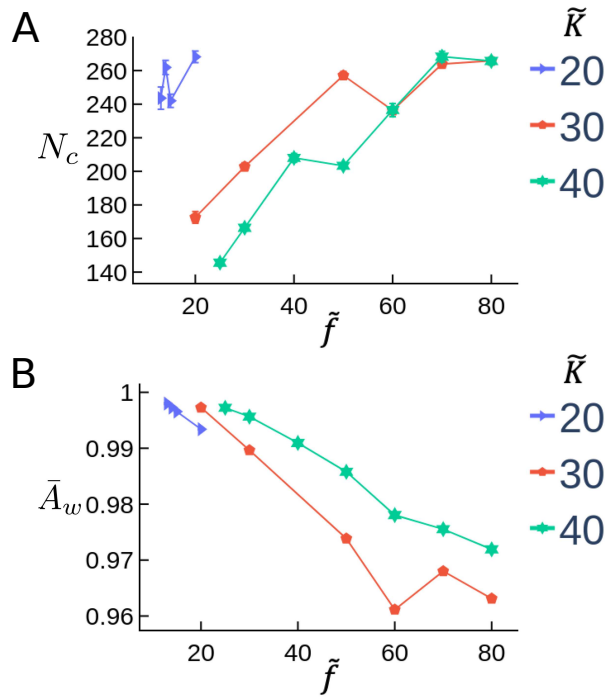}
  \caption{\label{fig:asphericity}Number of clusters and size-weighted average asphericity at different spring constants and driving forces in the cluster state. The data points are mostly for $\tilde{K}=30$ and $\tilde{K}=40$.  (A) The number of clusters $N_c$ increases as we increase the external driving force. We also observe that at the same external driving force, the number of clusters is larger with smaller spring constants. (B) The size-weighted average asphericity $\bar{A}_{w}$ quantifies how round the clusters are and gives more weight to larger clusters. Lower asphericity means the shape of the cluster is more spherically symmetric. The plot shows that the large clusters become more spherically symmetric as we increase the external driving force. At the same external driving force, the large clusters are more spherically symmetric in the systems with lower spring constants.}
\end{figure}

\subsection{Asphericity of the clusters}

\begin{figure*}
  \includegraphics[width=\linewidth]{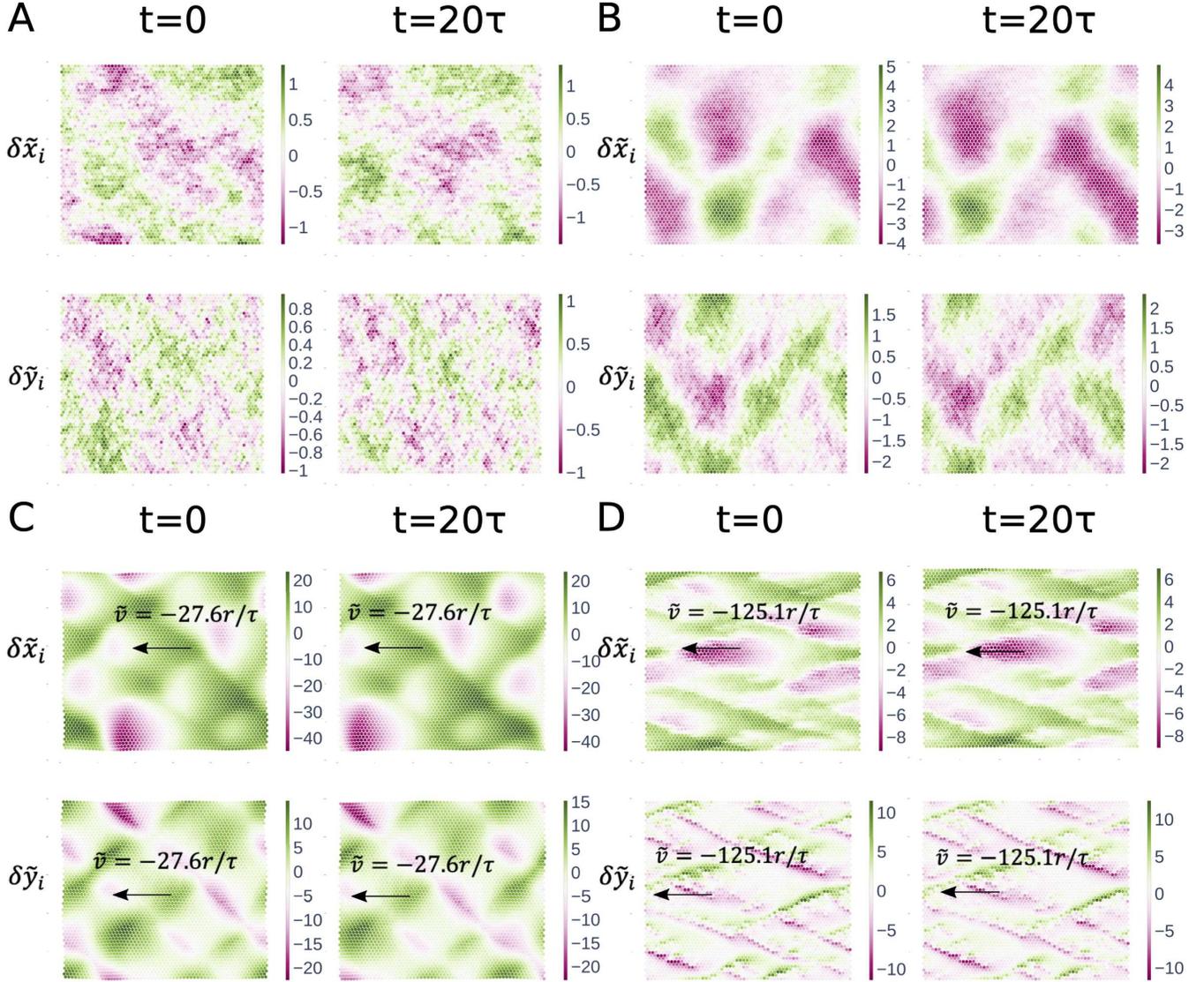}
  \caption{\label{fig:vibration}Vibrational motions of the large particles in the center of mass (COM) frame. In all figures, from left to right is time evolution, and the top two figures are the vibrational motions in the x-direction while the bottom two figures are the vibrational motions in the y-direction. The displacement vector fields are calculated in two consecutive microstates (snapshots) at $t$ and $t+20\tau$, where we set the first timestep $t=0$. The time interval $20\tau$ is chosen because it is large enough to observe the displacement of the pattern. The green color denotes positive values, which means vibrating in the direction of the axis, while the magnenta color means vibrating in the opposite direction of the axis. The color bars show the scale in the unit of small particles' radius $r$. (A) The dimensionless spring constant and driving force are $\tilde{K}=30$ and $\tilde{f}=10$, respectively. The system is in the lane state, and the vibrational motions of the large particles do not show a stable pattern. (B) The dimensionless spring constant and driving force are $\tilde{K}=30$ and $\tilde{f}=15$, respectively. The system is in the intermediate state. The vibrational motions of the large particles show patterns when time evolves, but the patterns are not stable (C) The dimensionless spring constant and driving force are $\tilde{K}=30$ and $\tilde{f}=20$, respectively. The system is in the cluster state, and stable patterns are observed. The wavepackets are moving at speed $\tilde{v}=-27.6r/\tau$. (D) The dimensionless spring constant and driving force are $\tilde{K}=30$ and $\tilde{f}=80$, respectively. The system is in the cluster state at a higher field, where we see stable patterns in the vibrational motions. The angle pattern is due to the separation between the clusters in the neighbor channels. We can see some constant horizontal separations between the wavepackets of displacement vectors, and they are defined as the length scales of the vibrational pattern. The wavepackets are moving at speed $\tilde{v}=-125.1r/\tau$.}
\end{figure*}
To quantify the evolution of the shapes of the clusters, we introduce the concept of asphericity. One cluster is a collection of mass points. The gyration tensor of the cluster is defined as
\begin{equation}
S_{mn}=\frac{1}{2N_{p}^2}\mathlarger{\mathlarger{\sum}}^{N_{p}}_{i=1}\mathlarger{\mathlarger{\sum}}^{N_{p}}_{j=1}({r_m}^{(i)}-{r_m}^{(j)})({r_n}^{(i)}-{r_n}^{(j)})    
\end{equation}
where $m, n$ denote the directions, $i, j$ denote the particle indices and $N_{p}$ denotes the number of particles in the cluster. This tensor describes the second moments of positions of a collection of particles. We can always find a coordination system where $S_{mn}$ is diagonal, 
\begin{equation}
    \boldsymbol{S}=\begin{bmatrix}
    \lambda^2_{a} & 0 \\
  0 & \lambda^2_{b}
  \end{bmatrix}
\end{equation}
where we assume $\lambda_a \leq \lambda_b$. The cluster asphericity is defined as $A=\frac{2(\lambda^4_{a}+\lambda^4_{b})}{(\lambda^2_{a}+\lambda^2_{b})^2}-1$. Under such definition, $A=1$ when the cluster is completely flat, i.e., $\lambda_{a}=0$. $A=0$ when the cluster is spherically symmetric, i.e., $\lambda_a=\lambda_b$. We observe that the shapes of small clusters do not change significantly when we increase the driving force since small clusters do not interact with the lattice as strongly as large clusters. Thus, we use the size-weighted average asphericity to quantify the change of shapes of clusters, giving more weight to larger clusters. The size-weighted average asphericity is defined as
\begin{equation}
\bar{A}_{w}=\frac{\mathlarger{\mathlarger{\sum}}^{N_{c}}_{i=1}{N_{p}}^{(i)}*A^{(i)}}{\mathlarger{\mathlarger{\sum}}^{N_{c}}_{i=1}{N_{p}}^{(i)}}
\end{equation}
where $N_c$ denotes the number of clusters, ${N_{p}}^{(i)}$ denotes the number of particles in the $ith$ cluster, and $A^{(i)}$ denotes the asphericity of the $ith$ cluster. FIG. \ref{fig:asphericity}A shows that the number of clusters $N_c$ increases as the external driving force increases and eventually saturates at high external fields. The average value of the number of clusters is calculated from 10 consecutive timeframes. The errorbar is calculated from the standard deviation of the number of clusters over 10 consecutive timeframes. The system is considered to reach the nonequilibrium steady state (NESS) if the standard deviation of the number of clusters is less than 0.1 of the average number of clusters, i.e., the number of clusters reaches a stable value. This criterion for the nonequilibrium steady state (NESS) also works for the lane state, since the number of clusters is well defined also in the lane state. In the lane state, the number of clusters equals the number of channels, which is 60 in our simulations. In FIG. \ref{fig:asphericity}B, we see that size-weighted average asphericity decreases when driving force increases. This means that the larger clusters become more spherically symmetric as we increase the driving force. It also shows that the larger clusters are more spherically symmetric in the systems with lower spring constants. This observation and the phase diagram can be understood better when we look at the vibrational motions of the large particles.

\subsection{Soliton-like vibrational motions of the large particles}

\begin{figure*}
  \includegraphics[width=\linewidth]{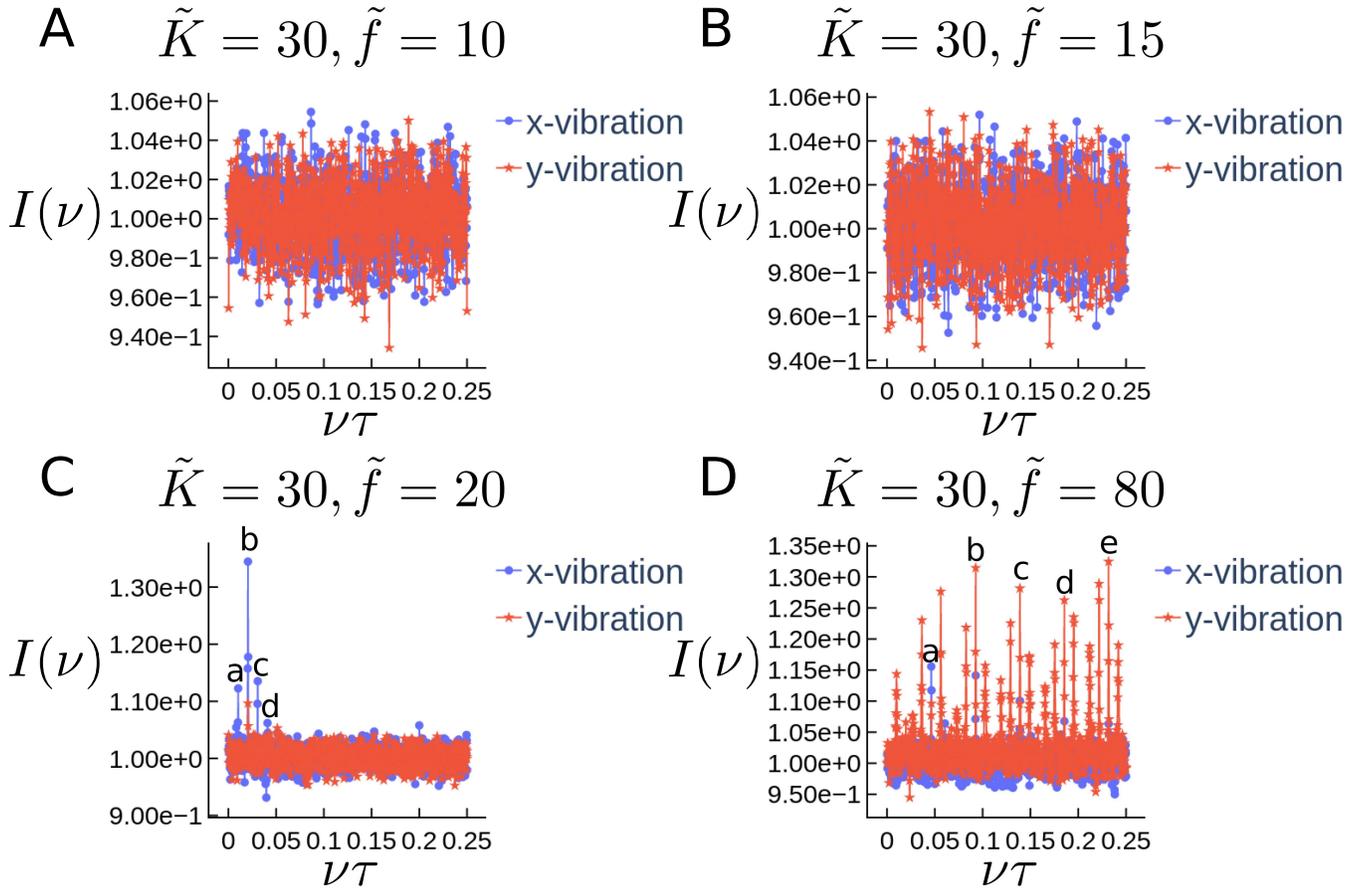}
  \caption{\label{fig:vibration_spectra}Vibrational spectra of large particles in the center of mass (COM) frame. In all figures, the blue lines are the spectra of the vibrational motion in the x-direction and the red lines are the spectra of the vibrational motion in the y direction. The horizontal axis is $\nu\tau$ because we express the frequency $\nu$ in the unit of $\frac{1}{\tau}$, where $\tau$ is the time unit in our simulations. (A) The dimensionless spring constant and driving force are $\tilde{K}=30$ and $\tilde{f}=10$, respectively. The system is in the lane state, and the vibrational spectra of the large particles do not have peaks. (B) The dimensionless spring constant and driving force are $\tilde{K}=30$ and $\tilde{f}=15$, respectively. The system is in the intermediate state, and the vibrational spectra of the large particles do not have peaks. (C) The dimensionless spring constant and driving force are $\tilde{K}=30$ and $\tilde{f}=20$, respectively. The system is in the cluster state, and the vibrational spectra of the large particles show several peaks. The four peaks $a$, $b$, $c$, $d$ marked in the graph correspond to $\nu\tau=0.0103,\ 0.0205,\ 0.0308,\ 0.0410$ respectively. They are all multiples of the frequency of the first peak $\nu_a\tau=0.0103$, which is the characteristic of a non-trigonometric periodic function with a period of $T=\frac{1}{\nu_a}=97.1\tau$. This period corresponds to the time the stable vibrational patterns travel over one box length in the x-direction. (D) The dimensionless spring constant and driving force are $\tilde{K}=30$ and $\tilde{f}=80$, respectively. The system is in the cluster state, and the vibrational spectra of the large particles show several peaks. The five peaks $a$, $b$, $c$, $d$, $e$ marked in the graph correspond to $\nu\tau=0.0463,\ 0.0928,\ 0.1390,\ 0.1855,\ 0.2318$ respectively. They are all multiples of the frequency of the first peak $\nu_a\tau=0.0463$, which is the characteristic of a non-trigonometric periodic function with a period of $T=\frac{1}{\nu_a}=21.6\tau$. This period corresponds to the time the stable vibrational patterns travel over one box length in the x-direction. There are many other peaks in the spectra, which correspond to the length scales of the vibrational pattern. }
\end{figure*}

When the system evolves from the lane state to the cluster state, we observe that the vibrational motions of the large particles show different behavior in different states. In the lane state, the vibrational amplitude is very small, and does not show stable patterns when time evolves (FIG. \ref{fig:vibration}A). In the intermediate state, we observe the correlation between the vibrational motions in different timesteps, but the pattern is not stable (FIG. \ref{fig:vibration}B). In the cluster state, we observe a stable pattern  (FIG. \ref{fig:vibration}C and D). In the COM frame, the large particles vibrate around their reference positions, and the small particles drift to the left at the average velocity $\tilde{\textbf{v}}=\textbf{v}_{small}-\textbf{v}_{large}$, where the $\textbf{v}_{small}$ and $\textbf{v}_{large}$ are the average drift velocities of the small and large particles in the lab frame. In the COM frame, the vibrational motions of the large particles in the cluster state satisfy the equation $\delta\tilde{\boldsymbol{R}}_{i}(t)=\delta\tilde{\boldsymbol{R}}_{i}(\tilde{\boldsymbol{R}}_{i0}-\tilde{\boldsymbol{v}}t)$, which means the wavepackets of the displacement vectors maintain their shapes while moving at the same velocity as the drift velocity of the small particles. This phenomenon is reminiscent of the soliton-like behavior observed in nonlinear dissipative systems \cite{Chetverikov2006,Malomed1996,Chetverikov2011}. From the vibrational spectra of large particles (FIG. \ref{fig:vibration_spectra}), we can quantitatively evaluate the transition. In the lane and intermediate state (FIG. \ref{fig:vibration_spectra}A and B), the vibrational spectra of large particles do not have peaks. In the cluster state (FIG. \ref{fig:vibration_spectra}C and D), the vibrational spectra show characteristics of non-trigonometric periodic functions. The period corresponds to the time the stable vibrational patterns travel over one box length in the x-direction in each condition. This feature shows that the vibrational motions of large particles have stable patterns that travel at constant velocity in the cluster state. In FIG. \ref{fig:vibration_spectra}D, we also observe many other peaks that correspond to the length scales of the vibrational pattern.  The displacement vectors of the large particles are correlated to the density distribution of the small particles. As the driving force increases, the large clusters have larger asphericity and tend to create more strain in the lattice. On the other hand, the larger the spring constant is, the harder the lattice can deform. That explains why the larger clusters are more spherically symmetric in the systems with lower spring constants. This indicates that the formation of clusters is the result of competition between the instability caused by the external driving force and the stiffness of the lattice. This physical picture is consistent with our observation in the phase diagram that the transition driving force increases as the spring constant increases.

\section{\label{sec:Conclusion}Conclusions}

In size-asymmetric charged colloidal compounds with springs attached between the large particles, we observe various steady states including a cluster state when the external driving force is higher than the lane state. In the cluster state, the small particles form clusters that travel at constant velocities with stable shapes and we observe soliton-like behavior in the vibrational motions of the large particles in the COM frame. As we increase the external field, the size-weighted average asphericity of clusters of small particles decreases, meaning that large clusters become more spherically symmetric. At the same time, it is more difficult for the lattice to deform with higher spring constants. Since no cluster state is observed when the large particles are frozen, this state is distinguished from the cluster formation in the active fluids \cite{Buttinoni2013, Zaccone2009,Kanehl2017,Zaccone2011}. The soliton-like vibrational motions are also different from the solitary wave observed in the nonlinear active lattices \cite{Chetverikov2006,Velarde2005} since our systems do not have active frictions and nonlinear springs. The cluster formation and soliton-like vibrational motions in our model systems are the result of the coupling between vibrational and fluid-like degrees of freedom under nonequilibrium conditions, which is similar to the solitons observed in the drainage of foams \cite{Koehler2000,Zhu2021}. In a driven binary lattice gas model \cite{Lavrentovich2021,Dickman2018}, phase separation with a long range order is also observed in the nonequilibrium steady state. In their model, the instability is induced by fluctuation correlations of the charge field, which could also be the origin of instability in our systems. It is important to note that the Debye length in our system is smaller than the length of the clusters since the system is on a surface in contact with an electrolyte (a reservoir containing ions), which provides electrostatic screening. In a system with long-range Coulombic interactions, the requirement of charge neutrality will forbid the formation of clusters. In a driven colloidal binary mixture with low volume fraction and long-range interactions, it was found that hydrodynamics play an important role in the self-assembly process \cite{Yuan2022}. In our systems, the hydrodynamics interactions are expected to be screened because the momentum will be transferred to the lattice of the large particles. However, hydrodynamics could still affect the morphology of the clusters in length scales of the order of the lattice constant, which can be a subject of future investigation. It should be noted that our systems have a very large parametric space, partly due to the size and charge asymmetries. We study extremely large size ratio and number ratio so that the small particles act like continuous flow over the size of the large particles. If the size ratio is not sufficiently large, we do not expect to observe the aggregation of the small particles. We also study a large overall density of the particles so that the scattering process between small and large particles is important. The phenomenon in this report is expected to be robust under such conditions, although we have not fully explored the large parametric space. Our work provides insight into the nonequilibrium organization behavior of size- and charge-asymmetric colloidal mixtures.
 
\vspace*{10mm}

\begin{acknowledgments}

We acknowledge Dr. Eleftherios Kyrkinis, Dr. Felipe Jim\'enez-\'Angeles, and Dr. Leticia L\'opez for their helpful discussions. We appreciate the valuable input from the reviewers. This work was supported by the Center for Bio-Inspired Energy Science (CBES), an Energy Frontier Research Center funded by the US Department of Energy (DOE) Office of Basic Energy Sciences (DE-SC0000989). We thank the computational support of the Fairchild Foundation and the Center for Computation and Theory of Soft Materials at Robert R. McCormick School of Engineering and Applied Science, Northwestern University.

\end{acknowledgments}

\nocite{*}

\end{document}